\documentclass[a4paper]{jpconf}
\usepackage{graphicx}
\begin{document}
\title{Transition barrier at a first-order phase transition in the canonical and microcanonical ensemble}

\author{Wolfhard Janke$^1$, Philipp Schierz$^1$ and Johannes Zierenberg$^{1,2,3}$}
\address{$^1$ Institut f\"ur Theoretische Physik, Universit\"at Leipzig, Postfach 100\,920, 04009 Leipzig, Germany }
\address{$^2$ Max Planck Institute for Dynamics and Self-Organization, Am Fassberg 17, 37077 G{\"o}ttingen, Germany}
\address{$^3$ Bernstein Center for Computational Neuroscience, Am Fassberg 17, 37077 G{\"o}ttingen, Germany}
\ead{wolfhard.janke@itp.uni-leipzig.de}

\begin{abstract}
  We compare the transition barrier that accompanies a first-order phase
  transition in the canonical and microcanonical ensemble. This is directly
  encoded in the probability distributions of standard Metropolis Monte Carlo 
  simulations and a proper microcanonical sampling technique. For the example 
  of droplet formation, we find that in both ensembles the transition barrier 
  scales as expected but that the barrier is much smaller in the microcanonical ensemble.
  In addition its growth with system size is weaker which will enhance this 
  difference for larger systems. We provide an intuitive 
  physical explanation for this observation.
\end{abstract}

\section{Introduction}
Many relevant transitions in nature fall into  the class of first-order phase
transitions, where directly at the transition point two or more phases coexist,
separated by highly suppressed transition states. Prime examples of first-order
transitions include the gas-liquid transition or the change of magnetic order
under variation of an external field. 

The formation of a droplet in a supersaturated gas is also a first-order phase
transition, where instead of coexistence of two pure phases (e.g., gas and
liquid) one observes coexistence between the homogeneous gas phase and 
a phase with a single droplet in equilibrium with the surrounding vapor. 
While the basic properties of this nucleation process have been thoroughly 
discussed~\cite{volmer1925,becker1935,feder1966,oxtoby1992,kashchiev2000}, many puzzling details
remain open. In fact, nucleation rates predicted by simulations and measured 
in experiments still do not match. Classical nucleation theory connects the 
rate $R$ of droplet formation with the free-energy barrier $\Delta F$:
%
  $R = \kappa e^{-\beta \Delta F}$.
%
The kinetic prefactor $\kappa$ includes the kinetic details of the nucleation
process and the free-energy barrier may be related to the suppression in the
probability distribution of a suitable reaction coordinate. Both may be, in
principle, computed from Monte Carlo simulations~\cite{auer2001,
zierenberg2017NatComm}. This notion is commonly adapted from the canonical
ensemble and we hence refer to the free-energy barrier $\beta\Delta F$ as a 
general transition barrier $B$ in the following. 

In case of the condensation-evaporation transition both the droplet size and the
potential energy are suitable reaction coordinates. We focus here on the
potential-energy probability distribution $P(E_p)$ and immediately notice that
this strongly depends on the thermodynamic ensemble. If the system is in a heat bath (canonical
ensemble) there is an exchange of energy with the surrounding; if the system is
isolated (microcanonical ensemble) then there is no energy flux. Of course, in
the latter case the system may transfer kinetic to potential energy and vice versa, which
allows for a well-defined potential-energy probability distribution.

In the following, we will discuss the effect of the ensemble on the transition
barrier of a general first-order phase transition for which the energy is a
suitable reaction coordinate. In Sec.~\ref{secMethod} we describe a proper
microcanonical sampling technique that allows us to obtain the potential-energy
probability distribution in both ensembles. Combined with multi-histogram
reweighting techniques, we get direct access to the suppression of transition
states -- the transition barrier. The finite-size scaling is discussed in
Sec.~\ref{secResults} followed by an intuitive explanation of the somewhat
surprising results. We finish with our
conclusions in Sec.~\ref{secConclusions}.

\section{Method}
\label{secMethod}
We employ Monte Carlo simulations in the ``real'' microcanonical
ensemble~\cite{calvo2000,martinmayor2007,schierz2015,schierz2016,davis2017}, referring to the
conservation of total energy $E$ which is the standard textbook 
definition of the
microcanonical ensemble. We emphasize this since most previous
applications of the microcanonical ensemble in Monte Carlo
simulations have focused on the conservation of potential energy $E_p$. 
The reason for this will become clear when we briefly discuss the method
in the following. 
In addition to the conservation of the total energy $E$, we further fix the particle number
$N$ and the volume $V$. This defines the NVE ensemble with the partition
function 
$\Omega(E) = \int\int \mathcal{D}x~\mathcal{D}p~\delta(E-[E_p(x)+E_k(p)])$,
where $\mathcal{D}x$ denotes the integration over state space and $\mathcal{D}p$
over momentum space. Integrating out the momentum degrees of freedom, which enter the
kinetic energy $E_k=\sum_{i}p_i^2/2m$,
this can be reduced to the (restricted) potential-energy space. For $N$
particles in three dimensions one obtains~\cite{calvo2000}
\begin{equation}
  \Omega(E)=\frac{(2\pi m)^{\frac{3N}{2}}}{\Gamma(\frac{3N}{2})}\int_{-\infty}^{\infty} dE_p
  \hat{\Omega}(E_p) (E-E_p)^{\frac{3N-2}{2}}\Theta(E-E_p),
  \label{eqGamma}
\end{equation}
where $\Gamma(n)$ is the Gamma function, $\hat\Omega(E_p)$ is the
(conformational) density of states and $\Theta\left(E-E_p \right)$ is the
Heaviside step function reflecting the constraint $E_p \le E$. 

We can thus sample the microcanonical phase space by generating a Markov chain
according to the weight
\begin{equation}
 W_{\rm{NVE}}(E_p)=\left(E-E_p\right)^{\frac{3N-2}{2}}\Theta\left(E-E_p \right),
\end{equation}
where of course we have to start from a potential energy $E_p<E$ since $E_p>E$
has zero probability in this ensemble. The usual Metropolis acceptance probability 
for a proposed move from (micro)state $A$ to $B$ is then naturally adapted 
to the NVE ensemble:
\begin{equation}\label{eqImportanceSampling}
 P_{\rm{acc}}\left(A\rightarrow B\right)=\min\left\{1,W_{\rm{NVE}}(E_p^B)/W_{\rm{NVE}}(E_p^A)\right\}.
\end{equation}
Here one clearly sees the difference to fixing the potential energy in a
conformational microcanonical ensemble. While this might be a natural ensemble
for spin systems, where a kinetic contribution is not properly defined (but may
be exploited for numerical purposes~\cite{martinmayor2007}), it is an incomplete
ensemble for general systems in soft condensed matter. Of course, a standard
Monte Carlo approach in the canonical ensemble makes use of this reduction. This
is valid for all canonical expectation values that are independent of the 
kinetic
contributions. However, if one considers quantities that are derived from energy
probability distributions, e.g., the free-energy barrier, the contributions of
the kinetic energy turn out to play a role~\cite{zierenberg2017NatComm}.

The importance sampling
defined in (\ref{eqImportanceSampling}) can be combined with a replica-exchange
scheme, where parallel simulations at different total energies exchange their
configurations with the probability
\begin{equation}
 P_{\rm{exc}}\left(A\leftrightarrow B\right)=\min\left\{1,\frac{W_{\rm{NVE}^B}(E_p^A)W_{\rm{NVE}^A}(E_p^B)}{W_{\rm{NVE}^B}(E_p^B)W_{\rm{NVE}^A}(E_p^A)}\right\}.
\end{equation}
Afterwards, NVE WHAM can be applied to estimate the density of states
$\hat\Omega(E_p)$~\cite{schierz2015,Ferrenberg1989,kumar1992,kim2011}. 
This is
an ensemble-independent property of the simulated system and allows one to
estimate observables in other ensembles, e.g., the canonical NVT ensemble
\cite{schierz2015,escobedo2006}.


\newcommand{\eqhP}{P^{\rm{eqh}}}
\section{Results}
\label{secResults}

\begin{figure}
  \centering
  \includegraphics[width=0.45\textwidth]{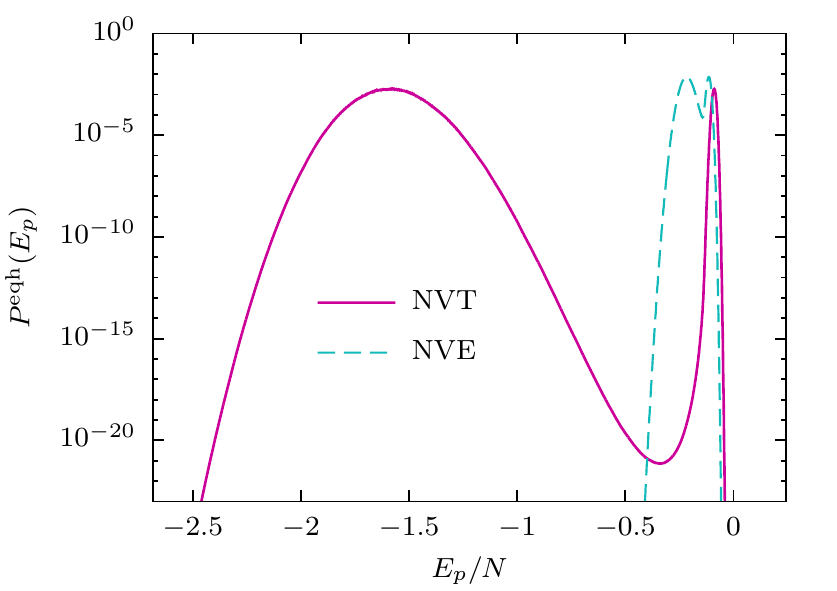}
  \includegraphics[width=0.45\textwidth]{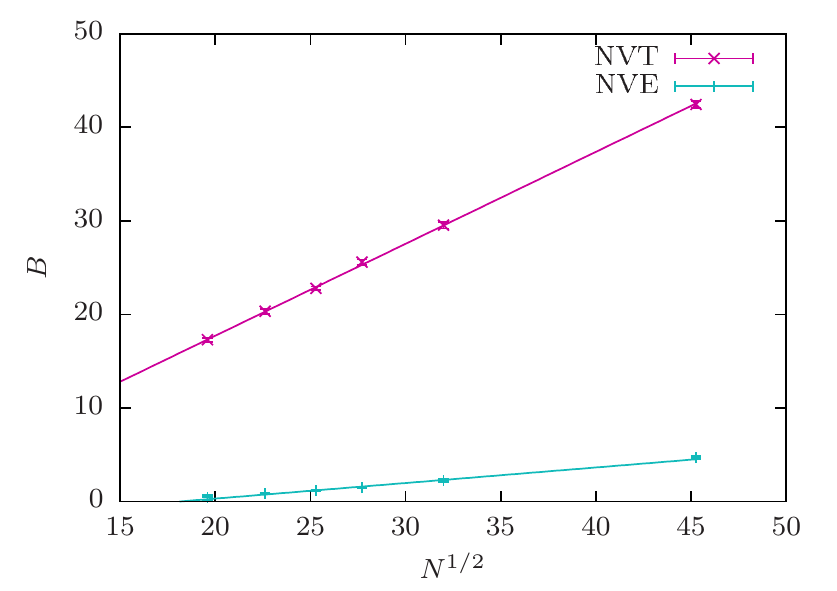}
  \caption{%
    (left) Probability distributions of the potential energy in the canonical (NVT) and
    microcanonical (NVE) ensemble for 2048 Lennard-Jones particles at fixed
    density $\rho=10^{-2}$. The distributions are obtained at the equal-height
    temperature $T_{\rm{eqh}}=0.6175$ and equal-height total energy
    $E_{\rm{eqh}}/N=0.6849$, respectively.
    (right) The free-energy barrier of the condensation-evaporation transition in a
    Lennard-Jones system for the (conformational) NVT and NVE ensembles. This is
    directly related to the sampling barrier in the respective importance
    sampling schemes.
    \label{figProbDist}
  }
\end{figure}

As an illustrative example we show results for the 12--6 Lennard-Jones particle
system where the particles $i$ and $j$ interact via the potential 
$V_{\rm LJ}(r_{ij}) = 4 \epsilon \left[(\sigma/r_{ij})^{12}-(\sigma/r_{ij})^6\right]$, 
cutoff and shifted at $r_{ij}=2.5\sigma$, with $\epsilon=1$ setting the energy 
(or temperature) and $\sigma=2^{-1/6}$ setting the length scale of the system. We
consider $N$ particles in a cubic box of volume $V$ with periodic boundary
conditions and fix the density to $\rho = N/V = 0.01$.
To update the particle positions, we
apply short-range displacement and long-range jump
proposals, with symmetric selection probabilities each. 
The combination of replica-exchange NVE sampling and NVE WHAM allows us to
directly estimate the potential-energy probability distribution in both the
canonical ensemble, $P_{\rm{NVT}}(E_p)\propto \hat\Omega(E_p) \exp(-E_p/k_BT)$, and
microcanonical ensemble, $P_{\rm{NVE}}(E_p)\propto \hat\Omega(E_p) W_{\rm{NVE}}(E_p)$.
Both show, as expected, a double peak at the condensation-evaporation
transition, see Fig.~\ref{figProbDist} (left). In the canonical ensemble, we
have to vary the temperature to obtain a wide distribution with pronounced peaks of equal
height at $T_{\rm{eqh}}=0.6175$. In the microcanonical ensemble, we vary the
total energy to obtain a comparably narrow double-peak distribution at
$E_{\rm{eqh}}/N=0.6849$.

In addition to being much narrower, we also notice that the suppression of
transition states in the microcanonical probability distribution is much lower
than in the canonical distribution. This corresponds to a much lower barrier
$B_{\rm{NVE}}\ll B_{\rm{NVT}}$, each defined by
\begin{equation}
 B=\ln \left[{\eqhP\left(E_p^{\pm}\right)}/{\eqhP\left(E_p^{0}\right)}\right],
\label{Bar_def}
\end{equation}
where $\eqhP (E_p)$ is the (equal-height) potential-energy distribution in the
respective ensemble, $E_p^{\pm}$ refers to the location of the two maxima and
$E_p^{0}$ to the location of the minimum in between. Figure \ref{figProbDist} (right) shows
the canonical and microcanonical barriers for system sizes $N\in\{384, 512, 640,
768, 1024, 2048\}$. The microcanonical barrier is always below the canonical
barrier. In fact, the suppression is several orders of magnitude smaller as
best illustrated for the $N=2048$ system where $B_{\rm{NVT}}=42.4$ and
$B_{\rm{NVE}}=4.7$ -- recall that this measures the logarithm of the suppression.
In this case, importance sampling in the NVE ensemble is of the order of
$10^{16}$ times more efficient than in the canonical counterpart for this
transition. This observation can be derived from the ensemble
weights~\cite{schierz2016} and thus is generally true for any first-order phase
transition with suppressed transition states in the energy probability
distribution. It can be further generalized to any properly defined probability
distribution and tailored ensemble weight.
%


From the canonical ensemble we know that the barrier is related to the surface
of the interface that separates the coexisting phases, here the liquid droplet
from the surrounding gas. We expect that the barrier is thus proportional to the
surface of the droplet, $B\propto\partial V_D$. The surface depends on the
droplet volume, $\partial V_D \propto V_D^{2/3}$. Due to the interplay of energy
minimization inside the droplet and entropy maximization in the surrounding gas,
the droplet volume is expected to grow with system size as $V_D \propto
V^{3/4}$~\cite{biskup2002, biskup2003, neuhaus2003, binder2003}, or at fixed
density as $V_D \propto N^{3/4}$~\cite{zierenberg2015}. Introducing an effective
interfacial free energy $\tau$ and considering additional logarithmic
corrections~\cite{langer1967,ryu2010,nussbaumer2010,prestipino2012,prestipino2013} we
arrive at the finite-size scaling ansatz in the canonical
ensemble~\cite{zierenberg2017NatComm}
\begin{equation}
  B = \tau N^{1/2} - \alpha\ln N + c,
  \label{fssFreeEnergy}
\end{equation}
where $\alpha$ and $c$ are constants. 

In the microcanonical ensemble, the setup directly at the transition energy is
very similar to the canonical ensemble: A fixed volume and particle number,
while the system switches between a homogeneous gas and a droplet in coexistence
with the surrounding vapor. In fact, the conservation of total energy introduces
a potential-energy reservoir from which the system can take potential energy to
form a droplet or where it can store it to form a gas. In a limited scope, this
is locally comparable to the canonical ensemble. We hence assume the same
dependence of the barrier on the droplet surface, which trivially scales with
the droplet volume. The relation between droplet size and system size is as well
assumed to be consistent with the canonical ensemble, assuming the energy and 
entropy arguments to be transferable. We thus assume the same scaling ansatz
(\ref{fssFreeEnergy}).

Figure \ref{figProbDist} (right) shows the canonical and microcanonical barriers
which satisfy to leading order the scaling ansatz (\ref{fssFreeEnergy}). For simplicity, we
only consider a few  system sizes and restrict fits to the 
effective behavior $B=\tau N^{1/2}+c$, which yields
$\tau_{\rm NVT}=0.98(2)$ and $\tau_{\rm NVE}=0.166(8)$, with goodness-of-fit
parameters  $Q\approx0.79$ and $Q\approx0.18$, respectively. Please note that we
here focus on the conformational canonical ensemble because of the direct
relation to the Monte Carlo sampling methods. Working with the full canonical
ensemble, however, only shifts the line but leaves the $\tau$ estimate
invariant~\cite{zierenberg2017NatComm}.

Let us finally give a physical picture that heuristically explains the
difference of the sampling behavior in the NVT and NVE ensembles for the droplet
condensation-evaporation transition. In the NVT ensemble, the system is in a
heat bath, which ensures that there is always energy available to be added to
the system. The result is a constant transition probability to higher
potential-energy configurations with the same energy step $\Delta E_p$
independent of the specific value of the potential energy. At the canonical
transition temperature, this allows in principle to transform the coexisting
(possibly large) droplet at low potential energy into a gas with high potential
energy. The situation is different in the NVE ensemble, where a distinct amount
of energy is available and distributed in potential and kinetic energy. The
potential energy may only be raised by transferring kinetic energy with
decreasing probability until none is left. This puts an upper bound on the
reachable potential energies. The result is a decreasing transition probability
to higher potential-energy configurations which therefore depends on the current
potential energy and goes to zero when approaching the upper bound. 
At the point where the energy versus entropy competition evens out in the NVT
ensemble, the NVE ensemble will still be stuck in the droplet phase since higher
energies show a higher suppression.  Consequently, phase coexistence occurs in
the NVE ensemble at a smaller droplet size, which explains the smaller barrier.

\section{Conclusions}
\label{secConclusions}
We demonstrated on the example of the condensation-evaporation transition that the transition
barrier of a first-order phase transition is always smaller in the
microcanonical ensemble compared to the canonical ensemble. In fact, a
finite-size scaling analysis reveals that this effect drastically increases in
the limit of increasing system sizes. A physical reason for this is that in a
heat bath (canonical ensemble) the system can in principle increase its
potential energy infinitely, because there is an infinite reservoir of potential
energy -- the heat bath. In the microcanonical ensemble, the finite reservoir of
potential energy due to the constraint of a fixed total energy allows only very
narrow distributions with a reduced barrier~\cite{schierz2016}. These general
statements about the ensembles are not restricted to computational
considerations. Physical microcanonical signatures might be approximately
observed in astrophysics or extremely isolated systems on earth where the
barrier difference should show up as well.

\section*{Acknowledgments}
This work has been partially supported by
  the DFG (Grant No.\ JA 483/31-1),
  the Leipzig Graduate School ``BuildMoNa'',
  and the Deutsch-Franz\"osische Hochschule DFH-UFA (Grant No.\ CDFA-02-07).
The authors gratefully acknowledge the computing time provided by the John von
Neumann Institute for Computing (NIC) on the supercomputer JURECA at J\"ulich
Supercomputing Centre (JSC) under Grant No.\ HLZ24.
JZ received financial support from the German Ministry of Education
and Research (BMBF) via the Bernstein Center for Computational Neuroscience
(BCCN) G{\"o}ttingen under Grant No.~01GQ1005B.

\section*{References}

\end{document}